\documentclass[journal]{IEEEtran}
\usepackage{graphicx}
\usepackage{csquotes}
\usepackage[utf8]{inputenc}
\usepackage[T1]{fontenc}
\usepackage{mathptmx}
\usepackage{comment}
\usepackage{subfig}
\usepackage{amssymb}
\usepackage{url}
\usepackage[dvipsnames]{xcolor}

\ifCLASSINFOpdf
\else
\fi

\begin{document}
%
\title{TiAu TES 32$\times$32 pixel array: uniformity, thermal crosstalk and performance at different X-ray energies}
%
%
%

\author{Emanuele~Taralli, Matteo~D'Andrea, Luciano~Gottardi, Kenishiro~Nagayoshi, Marcel~Ridder, Sven~Visser, Martin~de~Wit, Davide~Vaccaro, Hiroki~Akamatsu, Kevin~Ravensberg, Ruud~Hoogeveen, Marcel~Bruijn  and~Jian-Rong~Gao
\thanks{E. Taralli, L. Gottardi, K. Nagayoshi, M. Ridder, S. Visser, M. de Wit, D. Vaccaro, H. Akamatsu, K. Ravensberg, R. Hoogeveen, M. Bruijn
        and J.R. Gao are with the Netherlands Institute for Space Research NWO-I/SRON, Utrecht, 3584 CA, The Nederland e-mail: (see e.taralli@sron.nl).}
\thanks{J.R. Gao are also with Faculty of Applied Science, Delft University of Technology, 2600 AA Delft, The Netherlands.}
\thanks{M. D'andrea is with INAF/IAPS Roma, Via del Fosso del Cavaliere 100, 00133 Roma, Italy}
\thanks{Manuscript received April 19, 2005; revised August 26, 2015.}}

\markboth{Journal of \LaTeX\ Class Files,~Vol.~14, No.~8, August~2015}%
{Shell \MakeLowercase{\textit{et al.}}: Bare Demo of IEEEtran.cls for IEEE Journals}

\maketitle

\begin{abstract}
Large format arrays of transition edge sensor (TES) are crucial for the next generation of X-ray space observatories. Such arrays are required to achieve an energy resolution of $\mathrm{\Delta}E~\textless$~3~eV full-width-half-maximum (FWHM) in the soft X-ray energy range. We are currently developing X-ray microcalorimeter arrays as a backup option for the X-IFU instrument on board of ATHENA space telescope, led by ESA and foreseen to be launched in 2031.
In this contribution, we report on the development and the characterization of a uniform 32$\times$32 pixel array with (length$\times
$width) 140$\times$30 $\mu$m$^2$ TiAu TESs, which have \textcolor{black}{a 2.3 $\mu$m} thick Au absorber for X-ray photons. The pixels have a typical normal resistance $R_\mathrm{n}$ = 121 m$\Omega$ and a critical temperature $T_\mathrm{c}\sim$ 90 mK. We performed extensive measurements on 60 pixels out of the array in order to show the uniformity of the array. We obtained an energy resolutions between 2.4 and 2.6 eV (FWHM) at 5.9 keV, measured in a single-pixel mode at AC bias frequencies ranging from 1 to 5 MHz, with a frequency domain multiplexing (FDM) readout system, which is developed at SRON/VTT. We also present the detector energy resolution at X-ray with different photon energies generated by a modulated external X-ray source from 1.45 keV up to 8.9 keV.
Multiplexing readout across several pixels has also been performed to evaluate the impact of the thermal crosstalk to the instrument's energy resolution budget requirement. \textcolor{black}{This value results in a derived requirement, for the first neighbour, that is less than 1$\times$10$^{-3}$ when considering the ratio between the amplitude of the crosstalk signal to an X-ray pulse (for example at 5.9 keV)}.
\end{abstract}


\IEEEpeerreviewmaketitle

\section{Introduction}

\IEEEPARstart{A}{THENA} \cite{athena1} is the second \enquote*{Large mission} of ESA's Cosmic Vision-programme to study astrophysical phenomena near black holes and galaxy clusters. \textcolor{black}{One of the two  scientific instruments on board is the X-ray Integral Field Unit (X-IFU) \cite{athena2}, a detector consisting of an array of over 3000 Transition Edge Sensor (TES) calorimeters, sensitive in 0.2-12 keV energy range, with 2.5 eV energy resolution below 7 keV.} SRON is currently developing X-ray transition edge sensor (TES) microcalorimeter array as a backup technology for X-IFU.\\ 
During the recent past, we have been focused in the design optimisation of our TiAu bilayer TESs. Number of mixed and uniform arrays, with different pixel designs, have been successfully fabricated \cite{pourya,ken}. Square pixels \cite{pourya,taralli1} and high aspect ratio devices \cite{taralli2,martin} have been fully characterised using our frequency domain multiplexing (FDM) readout system and a preliminary comparison of the performance, between same devices biased under AC and DC, have been also investigated \cite{taralli3}. Lately, uniformity of  a 32$\times$32 pixels array have been extensively studied and the results have been widely reported \cite{taralli4}.\\
However, besides the homogeneity, many other aspects play a crucial role in fabrication and in the delivery of large arrays. Proper thermalisation process, to avoid undesired phonons diffusing through the substrate after an X-ray has been detected, can have a major impact in the final evaluation of the instrument's energy resolution budget. Moreover the detector energy resolution needs to remain excellent for a large range of the energy spectrum.\\
In this work, we first resume the main results in terms of uniformity of critical temperature and energy resolution at 5.9 keV among the 60 pixels measured over the uniform 32$\times$32 array. Afterwards we present thermal crosstalk ratio for four closest neighbours and eventually we conclude with the evaluation of the energy resolution at X-ray with different photon energy using a modulated X-ray source (MXS).

\section{Experimental setup}
\begin{figure}
\centering
\includegraphics[width=0.8\linewidth, keepaspectratio]{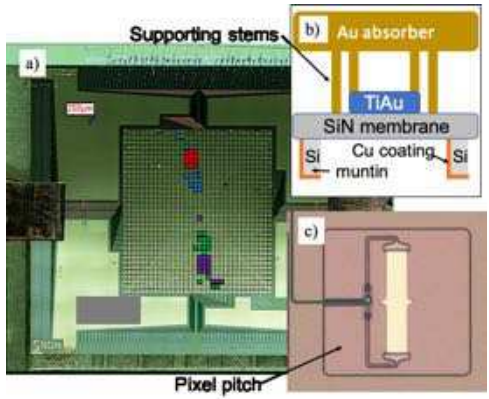}\\
\caption{\label{fig:array}(a) Top view of 32$\times$32 pixels array with the pixels measured during the 4 measurement runs: purple (Run1), green
(Run2), blue (Run3) and red (Run4). (b) Schematic of the single pixel (not to scale). Supporting stems hold the absorber above the TES which lays on the SiN membrane supported by the Si muntins coated with Cu. (c) Picture of a 140$\times$30 $\mu$m$^2$ TiAu TES .}
\end{figure}

We selected a total amount of 60 TESs \textcolor{black}{in order to investigate the performance from one side to the other} of a uniform 32$\times$32 pixels array and we measured them in 4 different measurement runs as highlighted by colours in Fig.~\ref{fig:array}a. All devices have the same aspect ratio (length$\times$width) of 140$\times$30 $\mu$m$^2$ (Fig.~\ref{fig:array}c) and the same bilayer thickness of Ti (35 nm) and Au (200 nm), which have been selected to show a sheet resistance of 25 m$\Omega$/$\square$ with superconducting transition temperature $T_\mathrm{c}\sim$90 mK. All absorbers have also the same size (240$\times$240 $\mu$m$^2$) and the same thickness (2.35 $\mu$m of Au). Each absorber is connected to the corners of the membrane by means of 4 supporting stems and the other two are thermally coupled to the centre of the TES, as shown in the schematic of a single pixel in Fig.~\ref{fig:array}b. \textcolor{black}{A 0.5 $\mu$m thick SiN membrane supports each pixel. They are located on a grid of 300 $\mu$m thick and 80 $\mu$m width Si muntins shaped by back-etching under each TES to ensure a weak thermal link to the bath.} The final pitch size is 250 $\mu$m. \textcolor{black}{Gold wire-bonds have been used to thermalize the whole chip to the bath connecting only the silicon area on the top part of the \textcolor{black}{chip}. To increase the thermal conductivity between the chip and the bath, 1 $\mu$m thick coating of Cu with Au capping layer has been added  only to the sides and bottom of the muntins in the Si grid region of the pixels, as shown in Fig.~\ref{fig:array}.} More details on the fabrication of this TES array can be found in \cite{ken}.

Frequency Domain Multiplexing (FDM) \cite{hiroki,davide} is used to read out these detectors. \textcolor{black}{Each detector is separated in frequency by placing it in series with LC resonators of specific frequencies. FDM readout applies a set of sinusoidal AC carriers in order to bias the TES detectors in their bias points. When a TES detector is hit by an X-ray photon, its own carrier is amplitude modulated.} \textcolor{black}{The frequency band assigned to each detector is selected in order to prevent the detectors from interacting with each other. In this way we are able to readout multiple TES pixels by means of one amplifier channel, which uses only one set of Superconducting QUantum Interference Devices (SQUID) current sensors. For this work we have used an 18-channels FDM readout system with a yield of $\sim$83\% (15 resonators available) and with bias frequencies between 1 MHz and to 5 MHz.} 

\section{Characterization}

\subsection{T$_c$ and $\Delta$E at 5.9 keV}
\begin{figure}
\centering
\includegraphics[width=0.8\linewidth, keepaspectratio]{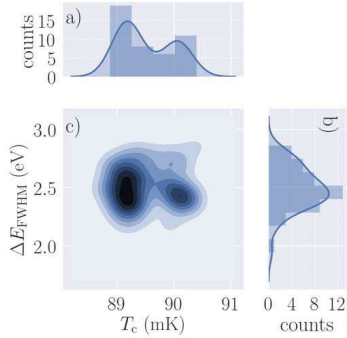}\\
\caption{\label{fig:ev_tc} (a) Histogram of the $T_\mathrm{c}$ measured. Two peaks correspond to the different $T_\mathrm{c}$ associated to the north and south quadrant of the chip, respectively. (b) Histogram of the energy resolution at 5.9 keV of the pixels measured in their best bias point. \textcolor{black}{The smoothed lines in both the histograms serve no other purpose than to guide the eye. (c)  Combined histogram where the contours define the most populated areas where detectors have similar $T_\mathrm{c}$ and $\mathrm{\Delta}E$.}}
\end{figure}

Critical temperature $T_\mathrm{c}$ and spectral energy resolution $\mathrm{\Delta}E$  are generally correlated by the following expression $\mathrm{\Delta}E = \sqrt{kT_\mathrm{c}^2C}\propto T_\mathrm{c}^{3/2}$ \textcolor{black}{where $k$ is the Boltzmann constant and $C$ is the detector heat capacity}. Low  $T_\mathrm{c}$ would certainly mean high resolving power and equally, uniform $T_\mathrm{c}$ over the whole array rather means uniform performance in detecting X-ray photon.\\
We have determined the critical temperature $T_\mathrm{c}$ over the array by measuring \textcolor{black}{the current-voltage characteristics ($IV$ curves)} of the pixels under test for different bath temperatures $T_\mathrm{bath}$ (from 50 mK up to 90 mK). Using these curves we calculate the dissipated electrical power $P_\mathrm{TES}$, for example at the minimum of the TES's $IV$ \textcolor{black}{along the phase transition}, at every bath temperature. Balancing the electrical and the thermal power dissipated by the TES, we are able to get and summarise all the $T_\mathrm{c}$ in the top histogram of Fig.~\ref{fig:ev_tc}a. The two peaks represent the critical temperatures  associated to the north and south quadrant of the array, respectively. It is worth noting  that all the pixels belonging to the same quadrant show a difference in the transition temperature less than 0.7 mK, while the total gradient from one side to the other of the array is less than 1.5 mK. We can consider the averaged $T_\mathrm{c} = 89.5\pm$0.4 mK. \textcolor{black}{This dispersion around the mean value might be explained considering some of the critical aspects in the whole fabrication process \cite{taralli4}.}\\
We have measured the energy resolution by exposing the TES array to a standard $^{55}$Fe source, providing Mn-K$\alpha$ X-rays at an energy of 5.9 keV with a count rate of approximately 1 \textcolor{black}{count per second (cps)} per pixel. \textcolor{black}{More information about the setup are reported in this recent work \cite{taralli4}.} Typically, we collect about 5000 X-ray events for each spectrum to get a statistical
error of about 0.15 - 0.18 eV for the reported energy resolution. Fig.~\ref{fig:ev_tc}b reports the histogram of all the energy resolutions measured for all the pixels under test in their best bias point, where the averaged energy resolution is 2.49$\pm$0.18 eV.\\
The 2-D histogram in Fig.~\ref{fig:ev_tc}c draws the main conclusion on the uniformity of the array highlighting that the large majority of the pixels show an  $\mathrm{\Delta}E_\mathrm{FWHM}$ between 2.4 and 2.6 eV in 1.5 mK variation of the critical temperature. 

\subsection{Thermal crosstalk}

\begin{figure}
\centering
\includegraphics[width=1.0\linewidth, keepaspectratio]{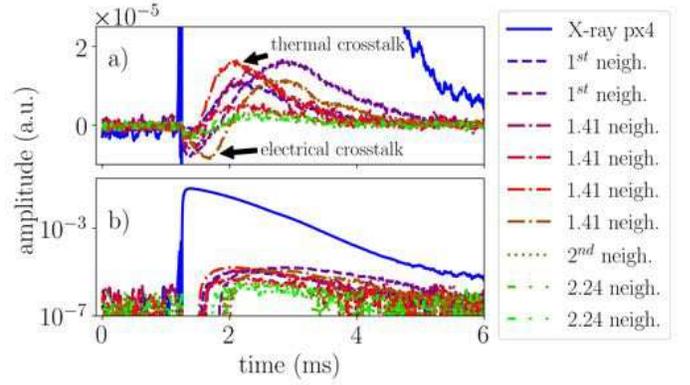}\\
\caption{\label{fig:pulses} Thermal pulses detected by victim pixels when X-ray photon (line) is absorbed by perpetrator pixel 4 in linear (a) and logarithmic scale (b). 1$^{st}$ neigh. \textcolor{black}{(dashed line)}, 1.41 neigh \textcolor{black}{(dash-dot line)}, 2$^{nd}$ neigh. \textcolor{black}{(dotted line)} and 2.24 neigh. \textcolor{black}{(dash-dot-dotted line)} listed in the legend are first, first diagonal, second and second diagonal neighbour of Px4, respectively.}
\end{figure}

\begin{figure}
\centering
\includegraphics[width=1\linewidth, keepaspectratio]{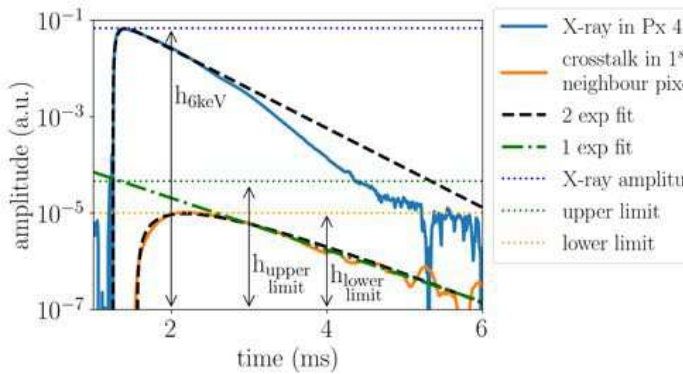}\\
\caption{\label{fig:up_low} Thermal pulse detected by the first neighbour when X-ray photon is absorbed by Px4. Black \textcolor{black}{dashed and green dot-dashed lines} define the double and single exponential fit functions respectively. Dotted lines define the three peak amplitudes used to evaluate the lower and upper limit of the thermal crosstalk.}
\end{figure}

\begin{figure*}
\centering
\includegraphics[width=0.45\linewidth, keepaspectratio]{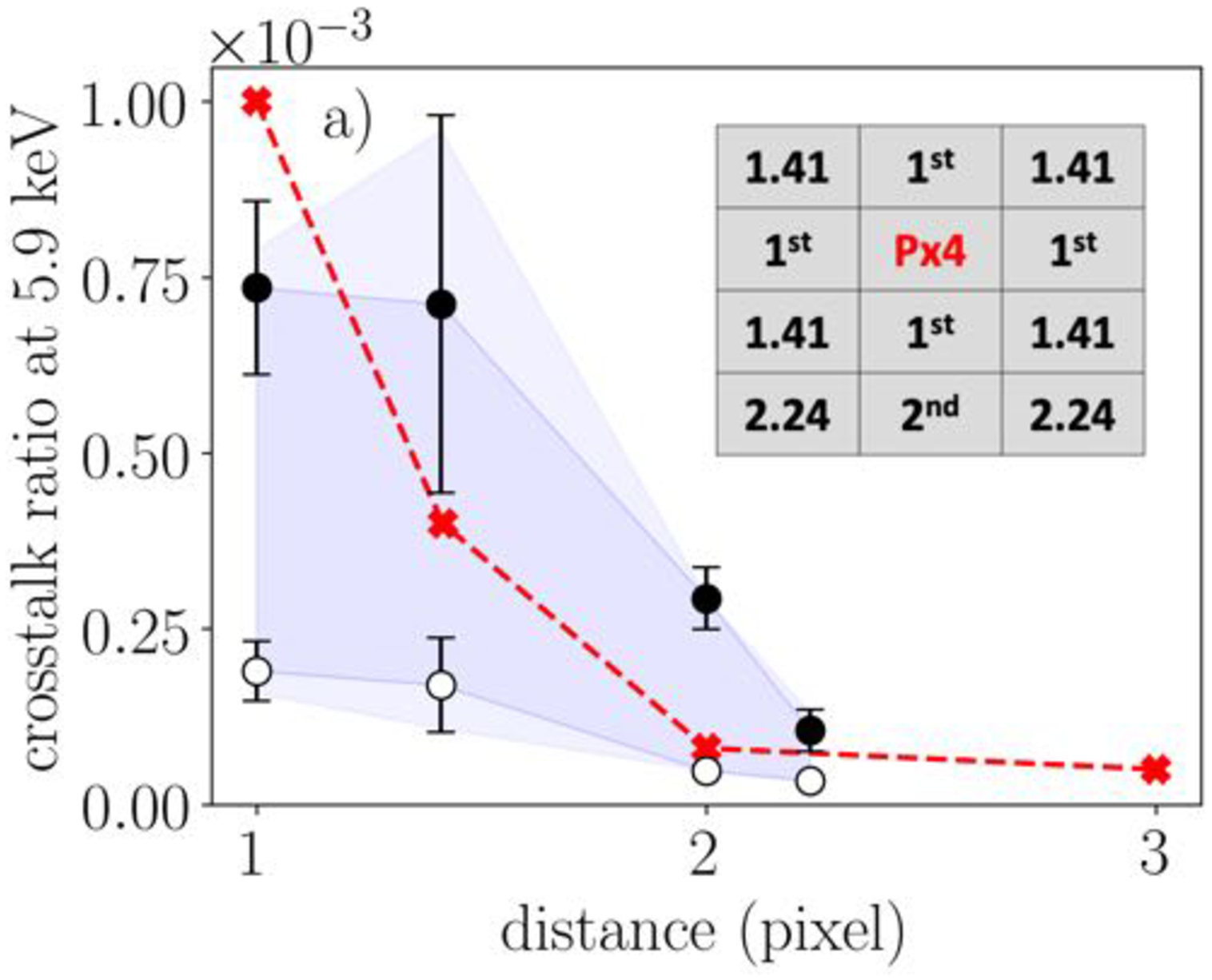}
\includegraphics[width=0.45\linewidth, keepaspectratio]{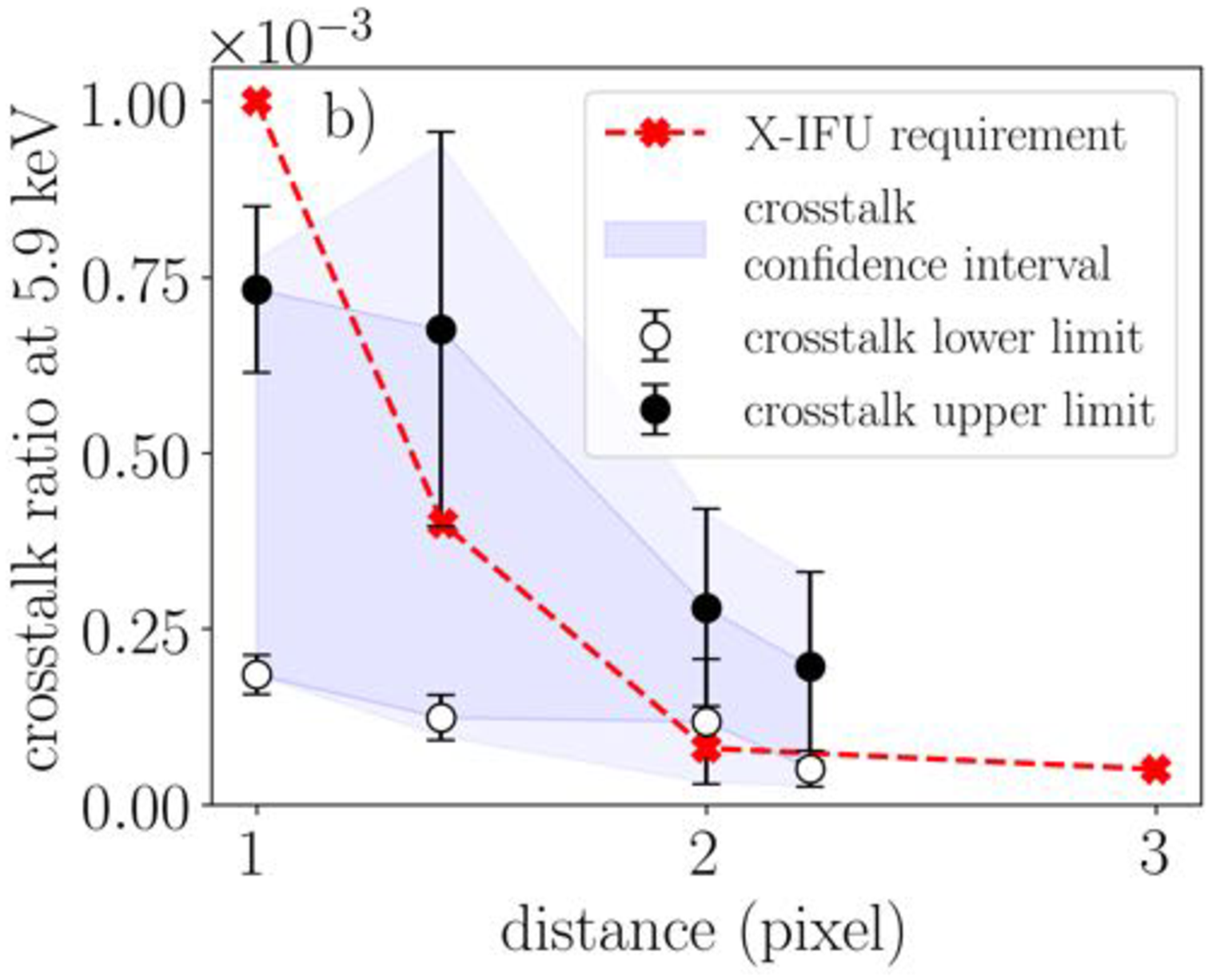}
\caption{\label{fig:crosstalk}Lower (open dots) and upper (closed dots) limit for the thermal crosstalk assessment \textcolor{black}{for pixel 4 (Px4) (a) and pixel 1 (Px1) (b)}. As an example, the inset in (a) shows the layout of the victim pixels around Px4. Grey area shows the region where we expect to find the ultimate thermal crosstalk of our kilo-pixel array. Red points refer to the X-IFU requirement.}
\end{figure*}

\textcolor{black}{Accurate thermalisation of large format and high-density arrays of microcalorimeters plays a crucial role in the minimisation of the thermal crosstalk between nearby pixels \cite{miniussi}}. \textcolor{black}{Facing this problem turns out to be crucial in order to meet the energy resolution requirements for the specific project} \cite{roland}. For instance, Athena has allocated 0.2 eV inside the instrument's energy resolution budget for the impact of the thermal crosstalk. It \textcolor{black}{derives} that the ratio between the amplitude of the crosstalk signal to an X-ray pulse (for example at 5.9 keV) is less than 1$\times$10$^{-3}$ (for the first neighbour), less than 4$\times$10$^{-4}$ (for the diagonal neighbour) and less than 8$\times$10$^{-5}$ (for the second nearest neighbour) \cite{roland}.
The main goal of such characterisation is to record any thermal pulse detected from neighbour pixels (named as victims) when a pixel (named as perpetrator) detects an X-ray photon.\\ 
In Run 4 (red pixels in Fig.~\ref{fig:array}a), we have performed thermal crosstalk measurement considering two perpetrators, e.g., \textcolor{black}{pixel 4 and pixel 1 (from now on named as Px4 and Px1)}, placed in the middle of the 12 pixels matrix. To minimise electrical crosstalk issues due to carrier leakage, all the detectors have been connected to \textcolor{black}{LC resonators} with bias frequency separated each other \textcolor{black}{by} at least 200 kHz. To further reduce this effect, we performed the measurement reading out only two pixels (one perpetrator and one victim) at the time. Fig.~\ref{fig:pulses} shows thermal pulses recorded by the victims pixels in linear and logarithmic scale respectively, when X-ray is detected for instance from Px4. We usually collect around 5000 X-ray counts from the perpetrator in order to get an averaged X-ray pulse. Each of these counts triggers the recording of the trace of the victim pixel. \textcolor{black}{Neglecting the traces corresponding to pileup event detected by the perpetrator}, we are able to define an averaged crosstalk pulse for the selected victim. Fig.~\ref{fig:pulses} reports the averaged crosstalk pulses for all the victims that have been considered. It also shows some pixels with a considerable negative spike just after the trigger induced by the X-ray detection. This is imputed to electrical crosstalk caused by the presence of common inductance both at the SQUID input and in the bias circuit and by mutual inductance in the LC-filter chip. This causes a scattering in the \textcolor{black}{starting point of the} thermal pulses. As a result, the electrical crosstalk effectively lower the peak amplitude of the thermal pulse adding an error in its evaluation.\\ 
In order to take into account the impact of this electrical crosstalk in the estimation of the thermal crosstalk, the analysis considers two types of fit: 1) a double exponentials fit function and 2) a single exponential fit function. 
The first one is used to extrapolate the peak amplitude of the pulse (dashed black line in Fig.~\ref{fig:up_low}). We have already said that this value is underestimated because the electrical crosstalk at the beginning of the pulse, as a matter of fact, reduces the actual peak of the thermal pulse. This provides a lower limit in the estimation of the pulse amplitude (orange dotted line in Fig.~\ref{fig:up_low}).
Since we do not have a reliable information on the thermal crosstalk pulse in its initial phase (as said, its amplitude is lowered), we use the second fit to extrapolate the peak value at the beginning of the thermal pulse considering the intersection of the fit function with the positive slope of the X-ray pulse (dash-dot green line in Fig.~\ref{fig:up_low}). This provides an upper limit in the estimation of the pulse amplitude (green dotted line in Fig.~\ref{fig:up_low}).\\
To assess the thermal crosstalk we finally consider the ratio between the amplitude of the thermal crosstalk signal from the victim to the X-ray pulse from the perpetrator, $crosstalk = h_\mathrm{victim}/h_\mathrm{6keV}$. In Fig.~\ref{fig:crosstalk} we plot the thermal crosstalk ratio measured for the perpetrator Px4 (Fig.~\ref{fig:crosstalk}a) and Px1 (Fig.~\ref{fig:crosstalk}b) as a function of  the neighbour pixels, respectively. The inset of Fig.~\ref{fig:crosstalk}a shows the layout of the victim pixels specifically for Px4 as perpetrator: first nearest neighbour pixel (distance 1), first diagonal neighbour pixel (distance 1.41), second nearest neighbour pixel (distance 2) and second diagonal neighbour pixel (distance 2.24). We have included open and closed dots as for the lower and upper limit of the crosstalk ratio, respectively. In this manner, we consider the impact of the electrical crosstalk defining a fair region where we expect to find the ultimate thermal crosstalk value. It is worth noting that we are well below the X-IFU requirement (red point in Fig.~\ref{fig:crosstalk}) for the first neighbour even considering the upper limit. Moreover, for reasons that still needs to be understood, the first diagonal neighbour (distance 1.41), in both the sequences considered, seems to be much more effected by the electrical crosstalk as indicated by the error bar reflecting a larger scattering in the peak amplitude estimation.  

\subsection{$\mathrm{\Delta}E$ at different X-ray energy photons}

\begin{figure}
\centering
\includegraphics[width=0.8\linewidth, keepaspectratio]{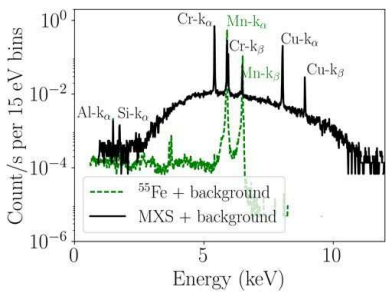}\\
\caption{\label{fig:spectra} Spectra \textcolor{black}{of the two sources used in this work to characterise the performance of this array: internal $^{55}$Fe source (green dashed line) and modulated X-ray source (black line).} The main energy lines are named in the graph.}
\end{figure}

\begin{figure}
\centering
\includegraphics[width=0.8\linewidth, keepaspectratio]{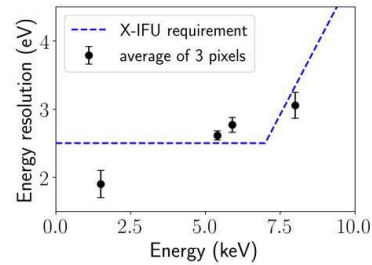}\\
\caption{\label{fig:ev_en} \textcolor{black} {Three pixels average energy resolution} as a function of energy \textcolor{black}{lines Al-k$\alpha$, Mn-k$\alpha$, Cr-k$\alpha$ and Cu-k$\alpha$} (black dots) compared to the X-IFU requirement as reference (blue dashed line). \textcolor{black}{Error bars refer to the standard deviation of the mean energy resolution obtained from three pixels.}}
\end{figure}

Large array of X-ray microcalorimers require to deliver high energy resolution over a large energy range in order to meet the instrument specification and consequently the science goal linked to the mission. Fig.~\ref{fig:spectra} shows the two spectra\textcolor{black}{, detected by a single TES,}  used in this work to characterise the performance of the array in terms of energy resolution. As already mentioned, the two Mn-k$\alpha$ and Mn-k$\beta$ (green dashed line in Fig.~\ref{fig:spectra}) \textcolor{black}{are obtained with an internal} $^{55}$Fe source, while the k$\alpha$ and k$\beta$ of Cr and Cu (black line in Fig.~\ref{fig:spectra}) \textcolor{black}{are obtained with} an external modulated X-ray source (MXS) \textcolor{black}{with a count rate of $\sim$2 cps}. Other two lines (Al-k$\alpha$ and Si-k$\alpha$) can be observed in the lower part of the spectra. They become visible only after having performed very long X-ray acquisition (more than 24 h). These lines are related to photons generated by the interaction between the X-ray photons coming from our internal $^{55}$Fe source with the Al parts of the setup (superconducting shield and foils to reduce the count rate) and the Si substrate of the TESs, respectively. More information about this characterization and the data analysis are provided in another paper which will be published in this special issue \cite{matteo}.\\ 
In this section we would like to give an overview of the energy resolution of the array as a function of the measured energy. We characterised three pixels Px1, Px8 and Px16 at different bias frequency 1, 2.7 and 4.6 MHz, respectively and the corresponding averaged energy resolution at four different energy lines \textcolor{black}{(Al-k$\alpha$, Mn-k$\alpha$, Cr-k$\alpha$ and Cu-k$\alpha$)} are reported in Fig.~\ref{fig:ev_en}. We would like to mention that the energy resolution can be properly evaluated only for the k$\alpha$ lines, the natural line shapes of which are well described \cite{kilburne} and have a decent number of counts as well. For this reason  Si-k$\alpha$, Mn-k$\beta$, Cr-k$\beta$ and Cu-k$\beta$ have been not included in the plot. \textcolor{black}{Notwithstanding a degradation of $\sim$ 4\% in the energy resolution between low and high bias frequency pixels \cite{taralli4}}, detector performance is very promising in the low (1.45 keV) and in the high (8 keV) part of the energy spectrum compered to the X-IFU requirement (blue dashed line). Improvements at 5.4 and 5.9 keV are foreseen with the new pixels design \textcolor{black}{currently under test.}

\section{Conclusion and next steps}
We are developing large uniform array of TiAu transition edge sensor microcalorimeters as backup option for X-IFU instrument on board of the Athena space mission. In this work we have presented the uniformity and the performance of a uniform 32$\times$32 pixels array, each of which is an TiAu TES with dimension (length$\times$width) 140$\times$30~$\mu$m$^2$ with a normal resistance $R_\mathrm{n}$~=~121~m$\Omega$. We have shown an averaged critical temperature $T_\mathrm{c}$~=~89.5$\pm$0.4~mK with a temperature spread from one side of the chip to the other $\Delta$T$_c$<1.5 mK. The large majority of the pixels have a $\mathrm{\Delta}E_\mathrm{FWHM}$ between 2.4 and 2.6 eV providing an averaged $\mathrm{\Delta}E_\mathrm{FWHM}$ = 2.49$\pm$0.18.\\
A first assessment of the thermal crosstalk has been performed, obtaining a value less than 1$\times$10$^{-3}$ for the first neighbour as required by Athena/X-IFU. The next step of this work is to perform a template for the electrical crosstalk pulse in order to remove its repercussion on the thermal pulse.\\
Energy resolution at different energy lines have been also evaluated finding an $\mathrm{\Delta}E_\mathrm{FWHM}$ less than 3 eV for X-ray energies up to 8 keV with a best $\mathrm{\Delta}E_\mathrm{FWHM}$ less then 2 eV at 1.45 keV. A next step is to characterize the large uniform arrays based on the new pixel design\textcolor{black}{, currently under test,} showing promising improvement at 5.9 keV in order to demonstrate a full array which meet the X-IFU requirement.

\section*{Acknowledgment}

This work is partly funded by European Space Agency (ESA) and coordinated with other European efforts under ESA CTP contract ITT AO/1-7947/14/NL/BW. It has also received funding from the European Union's Horizon 2020 Programme under the AHEAD (Activities for the High-Energy Astrophysics Domain) project with grant agreement number 654215.

\ifCLASSOPTIONcaptionsoff
  \newpage
\fi

\end{document}